# Analysis of Tidal Perturbations Due to Asymmetric Response of LARES 2 and LAGEOS

Xizhi Hu[1,4], Xiaodong Chen[2*,4] , Jianqiao Xu[1,4], Ignazio Ciufolini[3,5], Wei-Tou Ni[6,7], Antonio Paolozzi[5]

[1] State Key Laboratory of Precision Geodesy, Innovation Academy for Precision Measurement Science and Technology, CAS,Wuhan 430077, China

[2] Wuhan National Observation and Research Station for Geodesy, Innovation Academy for Precision Measurement Science and Technology, CAS,Wuhan 430077, China

[3] Wuhan Institute of Physics and Mathematics, Innovation Academy for Precision Measurement Science and Technology, Chinese Academy of Sciences, Wuhan 430071, China

[4] College of Earth and Planetary Sciences, University of Chinese Academy of Sciences, Beijing 100049, China

[5] Scuola di Ingegneria Aerospaziale, Sapienza Università di Roma, Rome 00138, Italy

[6] Lanzhou Center for Theoretical Physics and Key Laboratory of Theoretical Physics of Gansu Province, Lanzhou University, Lanzhou 730000, China

[7] International Center for Theoretical Physics Asia-Pacific, University of Chinese Academy of Sciences, Beijing 100190, China

* Corresponding author (email: chenxd@apm.ac.cn)

**Abstract** Earth tidal perturbations affecting laser-ranged satellites are critical for refining satellite orbital dynamics modeling, and their accurate computation represents a prerequisite for high-precision fundamental physical effects and geodetic investigations based on satellite orbit analysis. This study focuses on the tidal perturbations induced by the asymmetric responses of LARES 2 and LAGEOS on their orbital nodes and inclinations. Perturbations induced by a total of 402 (392 2nd- and 10 3rd-order) earth tide constituents on the two satellites were calculated, based on Kaula's orbital perturbation theory and Lagrange's planetary equations for satellites, considering the frequency dependence of Love numbers. The asymmetric characteristics of tidal perturbations between the two satellites were quantitatively analyzed. The minimum resolutions of orbital inclinations and nodes, used as screening thresholds for significant constituents, were derived from the RMS of overlapping orbit differences using orbital geometry and error propagation law. With these thresholds, 83 significant constituents were identified from the 402. The cumulative effect of the 319 minor constituents was further evaluated, and it was found that their total impact, from coherent superposition, noticeably exceeds the thresholds, thus becoming non-negligible. The results of this study provide accurate tidal perturbation parameters for LARES 2 and LAGEOS, and offer methodological references for the screening of Earth tide constituents in high-precision satellite orbital dynamics research, laying a foundation for subsequent studies on inverting geophysical parameters from satellite orbits and verifying fundamental physical effects, particularly the relativistic Lense-Thirring effect.

## 1. Introduction

Satellite Laser Ranging (SLR) to high-precision geodetic satellites represented by LARES 2 and LAGEOS is a core technical approach for refined research on satellite orbital dynamics, and the accurate modeling of orbital perturbation effects is the key to improving the precision of SLR orbit determination and application. Earth tides, as a typical time-varying gravitational perturbation source, induce the deformation of the Earth under the gravitational action of celestial bodies such as the Sun and the Moon, leading to the time-varying characteristics of the Earth's gravitational potential, which in turn causes periodic perturbations to the orbital elements of artificial satellites. For high-precision orbital dynamics research and application of LARES 2 and LAGEOS, the tiny errors of tidal perturbation modeling will be continuously amplified in the orbital evolution process, thus restricting the improvement of orbit determination accuracy and the reliability of subsequent geodetic and physical research based on satellite orbits. Therefore, the precise calculation of Earth tidal perturbations on the orbits of LARES 2 and LAGEOS, and the in-depth analysis of their perturbation response characteristics, are of important theoretical and practical significance for perfecting the satellite orbital dynamics model and improving the SLR orbit determination accuracy.

LARES 2 and LAGEOS are designed with a distinctive "butterfly" orbital configuration with suppulmentary orbital parameters for high-precision orbital dynamics research: their semi-major axes are nearly the same, the orbits are close to circular with near-zero eccentricities, and the orbital inclinations are highly supplementary—LARES 2 operates in a prograde orbit with inclination $i_{\text{LARES 2}}$=70.1615∘, while LAGEOS is in a retrograde orbit with $i_{\text{LAGEOS}}$=109.8469∘ (Ciufolini et al. 2023). This symmetric orbital design is intended to offset the classical orbital precession effects caused by the Earth's oblateness (e.g., the $J_2$ term) when combining the orbital signals of the two satellites, which provides a favorable orbital basis for high-precision research on satellite orbital perturbations. However, after offsetting the main Earth oblateness perturbation, the uncancelled tidal perturbation differences will be highlighted—the two satellites exhibit obvious asymmetric response characteristics to Earth tidal perturbations, and the tidal perturbation effects cannot be completely offset by the linear combination of orbital signals, which has become a key factor restricting the further improvement of the verification accuracy of Lense-Thirring effect. Clarifying the intrinsic mechanism of the asymmetric tidal perturbation response of the two satellites with symmetric orbital structures, and quantifying the perturbation differences of different Earth tide constituents on their key orbital elements (ascending node and inclination), is the core scientific problem to be solved in this study.

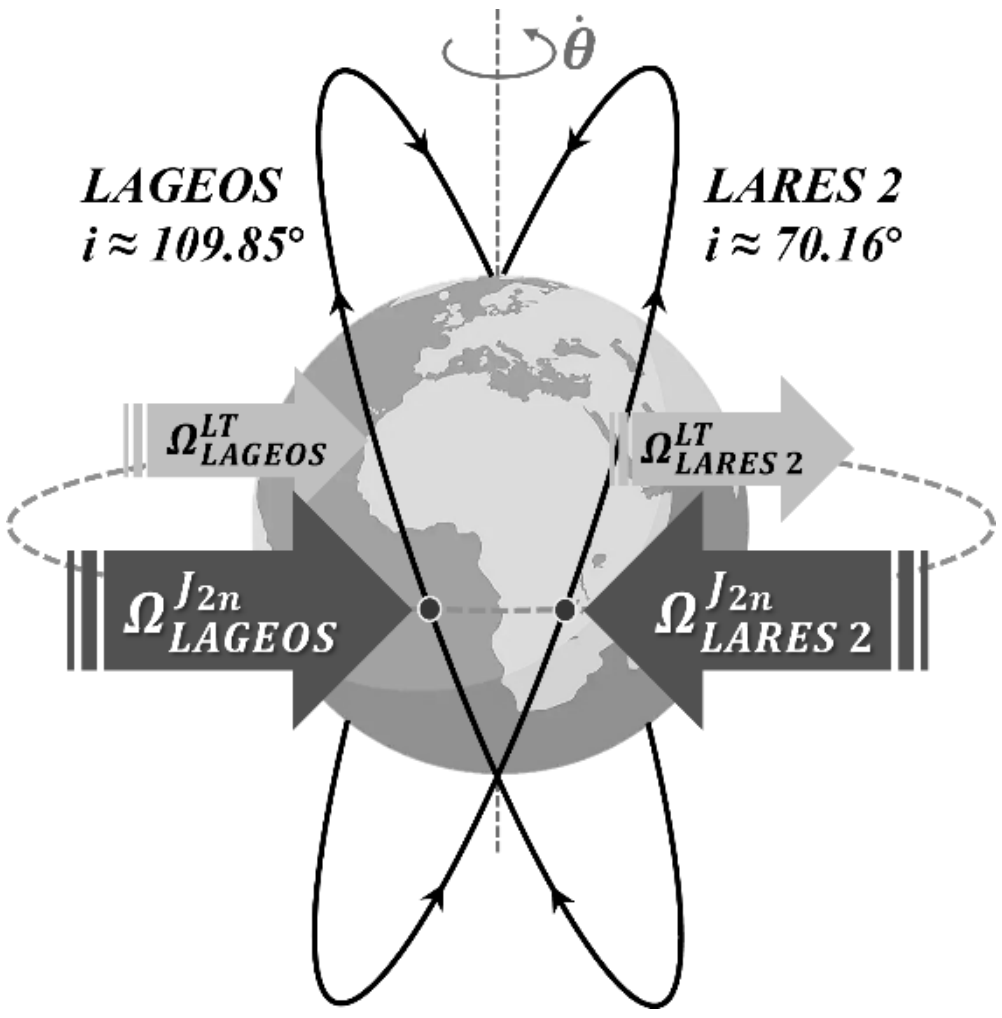

**Fig. 1** The "butterfly-shaped" orbital configuration of LAGEOS and LARES 2.

The fundamental reason for the asymmetric tidal perturbation response of LARES 2 and LAGEOS lies in the essential difference in the geodynamic response of prograde and retrograde orbits to Earth tides, which is reflected in the modulation effect of satellite orbital motion on the observed frequency of tidal perturbations. The observed frequency of tidal perturbations acting on satellites is jointly determined by the intrinsic frequency of tide-generating bodies (Sun, Moon) and the orbital motion characteristics of satellites, and contains the combination term $m(\dot{\Omega} - \dot{\theta})$ involving satellite nodal precession and Earth's rotation (see Equation 10). For LARES 2 and LAGEOS with prograde and retrograde orbital characteristics respectively, their nodal precession rates $\Omega$ have opposite signs; even for the same Earth tide constituent defined by the Doodson number (i.e., the same intrinsic tidal frequency $\sum_{i=1}^{6} 2\pi j_i/\gamma_i$), the existence of the $m(\dot{\Omega} - \dot{\theta})$ term leads to different tidal perturbation frequencies $\Gamma_{jlmp}$ experienced by the two satellites. The difference in perturbation frequency further results in the differences in the perturbation period and amplitude of the same tide constituent on the two satellites; in addition, the different initial positions of the two satellites determine the differences in the initial phase of perturbations, and the inability to align the phases makes it impossible to cancel each other out in the time domain. For harmonic perturbation signals, the difference in period makes the tidal perturbation effects of the two satellites unable to cancel each other through linear combination, and ultimately forms the asymmetric response characteristics of the two satellites to Earth tidal perturbations. To intuitively understand this mechanism, it can be analogized to two runners moving around a standard track at different constant speeds: the time interval of their encounter will be significantly different when they run in the same direction and in the opposite direction, which is similar to the difference in the modulation of tidal frequency by prograde and retrograde satellite orbits.

With the continuous advancement of Satellite Laser Ranging (SLR) technology, orbit determination accuracy has reached the sub-centimeter level, posing increasingly stringent requirements for the refinement of tidal perturbation modeling in high-precision orbital dynamics research of LARES and LAGEOS series satellites. In existing studies on Earth tidal perturbations of these satellites, scholars have conducted preliminary explorations on the screening of significant tidal constituents and perturbation calculation (Gurzadyan et al. 2017; Gurzadyan et al. 2025), successfully identifying partial significant Earth tide constituents that affect satellite orbits. However, this simplified modeling approach that only focuses on dominant tide constituents might no longer satisfy the future high-precision research needs, and these studies have not explicitly elaborated on the specific screening criteria adopted, cause the traditional screening method relying on a single constant amplitude threshold may have limitations. Specifically, the minor tidal constituents filtered out by the constant amplitude threshold often exhibit closely spaced frequency characteristics; their coherent superposition in the time domain is likely to produce a cumulative perturbation effect that cannot be ignored. This non-negligible cumulative effect will further introduce errors into satellite orbital modeling, thereby affecting the overall accuracy of orbital dynamics research.

The earth tidal perturbations on the nodes and inclinations of LARES 2 and LAGEOS were investigated in the study. We calculated the tidal perturbations and used the orbital element uncertainty from the inherent SLR orbit determination accuracy as the threshold to screen identifiable ones. Analysis of the filtered minor constituents showed their cumulative effect significantly exceeds preset thresholds, imposing new requirements for tidal perturbation screening methods. This study provided reliable tidal perturbation parameters for refined orbital modeling and improved SLR accuracy.

# 2. Tidal perturbation potential of an earth satellite

The Earth's spatial tidal potential $V_T$ is expressed via time variations of Stokes coefficients $C_{lm}$ and $S_{lm}$ in the spherical harmonic expansion of Earth's gravitational potential (Melchior 1983; Eanes 1983; Sanchez 1975). This arises because tide-generating bodies (primarily the Moon and Sun) lie far farther from Earth's center of mass than Earth's radius—thus, Earth's response to the external potential is linear, with the additional potential from Earth's shape deformation proportional to the external potential.

This study employs the linear perturbation formulation of Kaula (1966), using the tidal potential $V_T$ (minus) as a "disturbing function" to evaluate the changes in the orbital elements, specifically the right ascension of the ascending node (and the inclination), via Lagrange's planetary equations.

For the Earth's gravitational potential $V$:

$$V = -GM \sum_{l=2}^{\infty} \sum_{m=0}^{l} R^l V_{lm} \tag{1}$$

where $G$ is the gravitational constant, $M$ and $R$ are the mass and mean equatorial radius of Earth, respectively, and $V_{lm}$ are the components of the gravitational potential at different degrees l and orders m, with the form Kaula (1966):

$$V_{lm} = \frac{1}{a^{l+1}} \sum_{p=0}^{l} F_{lmp}(i) \sum_{q=-\infty}^{+\infty} G_{lpq}(e) \left[ \begin{pmatrix} C_{lm} \\ -S_{lm} \end{pmatrix}_{l-m \text{ odd}}^{l-m \text{ even}} \cos(\psi_{lmpq}) + \begin{pmatrix} S_{lm} \\ C_{lm} \end{pmatrix}_{l-m \text{ odd}}^{l-m \text{ even}} \sin(\psi_{lmpq}) \right] \tag{2}$$

where:

- $C_{lm}$, $S_{lm}$ are the normalized, dimensionless Stokes coefficients of Earth's gravitational potential;
- $F_{lmp}(i)$, $G_{lpq}(e)$: the inclination and eccentricity functions, respectively(Kaula 1961; 1964; 1966);
- $p$, $q$ are the expansion indices for the inclination and eccentricity functions, the "decomposition indices of the response end (satellite orbit)", which resolve the adaptation modes of orbital inclination and eccentricity to the gravitational perturbation.
- $\psi_{lmpq} = m(\Omega - \theta) + (l-2p)\omega + (l-2p+q)\mu$, where $\theta$ is the local sidereal time (Bertotti and Carpino 1989);
- $a$, $e$, $i$, $\Omega$, $\omega$, $\mu$ are the six Keplerian orbital elements of the satellite (semi-major axis, eccentricity, inclination, right ascension of the ascending node, argument of perigee, and mean anomaly), respectively.

Expressing $V_{lm}$ as a function of orbital elements in an inertial frame (not Earth's precessing spherical coordinates $(r, \phi, \lambda)$) enables direct "perturbation source-orbital response" modeling. This allows $V_{lm}$ to be used directly as the disturbing function in Lagrange's equations, avoiding complexity from reference frame dynamics (e.g., precession/nutation time-variability) or coordinate transformations—critical for efficient, accurate tidal effect analysis (e.g., in the calculation of $\Delta\Omega_{lmpq}$ and $\Delta i_{lmpq}$).

According to Kaula's theory, for an external body of mass $M^*$ with orbital elements $X^* = (a^*, e^*, i^*, \Omega^*, \omega^*, \mu^*)$ in the Earth's inertial coordinate system, the tidal potential component of degree l,m generated by it at a spatial point $(r, \phi, \lambda)$ can be written as (Efroimsky and Williams 2009):

$$V_{lm}^*(r, \phi, \lambda; X^*) = \frac{GM^*}{a^*} \left(\frac{R}{a^*}\right)^l \frac{(l-m)!}{(l+m)!} (2 - \delta_{0m}) P_{lm}(sin\phi) \times \sum_{p=0}^{l} F_{lmp}(i^*) \sum_{q=-\infty}^{+\infty} G_{lpq}(e^*) \begin{pmatrix} \cos \\ \sin \end{pmatrix}_{l-m \text{ odd}}^{l-m \text{ even}} (\psi_{lmpq}^* - m\lambda) \tag{3}$$

where $P_{lm}(sin\phi)$ is the normalized Legendre function, and $\psi_{lmpq}^*$ is determined by the Keplerian orbital elements of the tide-generating body:

$$\psi_{lmpq}^* = m(\Omega^* - \theta) + (l-2p)\omega^* + (l-2p+q)\mu^* \tag{4}$$

For Earth's response to the aforementioned tidal potential, linear tidal theory uses Love numbers $k_{lm}$ (Love 1926; Petit and Luzum 2010)—intrinsic elastic Earth parameters quantifying gravitational potential/deformation response to external forcing. At the satellite's position $(r, \phi, \lambda)$, the additional gravitational potential from Earth's tidal deformation is:

$$V_{\mathcal{T},\mathcal{S}} = -\sum_{l=2}^{+\infty} \sum_{m=0}^{l} k_{lm} \left(\frac{R}{r}\right)^{l+1} V_{lm}^{*}(r, \phi, \lambda; X^{*}) \tag{5}$$

The key point here is that the response potential $V_{\mathcal{T},\mathcal{S}}$ is proportional to the values of the tidal potential $V_{lm}^{*}$ at and outside Earth's surface $r \geq R$, and attenuates according to $\left(\frac{R}{r}\right)^{l+1}$, with the Love numbers $k_{lm}$ quantifying the magnitude of this response.

Since the additional gravitational potential is the deformation potential of Earth under the gravitational forces of the Sun and Moon, it must be referenced to Earth's center of mass, and satellite orbital perturbation analysis uses an inertial frame, Kaula's transformation is applied again to expand $P_{lm}(\sin\phi)$ using $F_{lmp}(i)$ and $G_{lpq}(e)$. This eliminates $\phi$, $\lambda$ and converts $V_{T,S}$ to the geocentric inertial frame. After index rearrangement, the disturbing function becomes:

$$V_{\mathcal{T},\mathcal{S}} = -\frac{GM^{*}}{a^{*}} \sum_{l=2}^{+\infty} \sum_{m=0}^{l} k_{lm} \left(\frac{R}{a}\right)^{l+1} \left(\frac{R}{a^{*}}\right)^{l} \frac{(l-m)!}{(l+m)!} (2-\delta_{0m}) \sum_{h=0}^{l} F_{lmh}(i^{*}) \sum_{j=-\infty}^{+\infty} G_{lhj}(e^{*}) \sum_{p=0}^{l} F_{lmp}(i) \times \sum_{q=-\infty}^{+\infty} G_{lpq}(e) \cos\left(\psi_{lmhj}^{*} + \psi_{lmpq} - \epsilon_{lmhj}\right) \tag{6}$$

where $\varepsilon_{lmhj} = \dot{\psi}_{lmhj}^{*} \Delta t_{lmhj}$, which represents the linear variation of the solar-lunar phase with time. $\dot{\psi}_{lmhj}^{*}$ is the angular frequency of the tidal component, which is the first derivative of equation (4) with respect to time (denoted later by $\sigma$ to distinguish from the satellite's angular frequency); $\Delta t_{lmhj}$ is the time difference from Earth's interior anelasticity. Furthermore, $\psi_{lmpq} = m(\Omega - \theta) + (l-2p)\omega + (l-2p+q)\mu$, where the physical meaning of the symbols here are similar to those in equation (4), except that it refers to the perturbed body, i.e., the artificial satellite. The expressions for the inclination functions $F_{lmp}(i)$ and eccentricity functions $G_{lpq}(e)$ of various degrees and orders used in this paper are shown in Table 1. The expressions for$F_{lmp}(i)$ and $G_{lpq}(e)$ differ for distinct values of $l$, $m$, $p$, and $q$. Specifically, the expressions $F_{201}(i) = \frac{3}{4}\sin^{2} i - \frac{1}{2}$, $F_{211}(i) = -\frac{3}{2}\sin i \cos i$, $F_{211}(i) = \frac{3}{2}\sin^{2} i$, and $G_{210}(e) = \frac{1}{\sqrt{(1-e^{2})^{3}}}$ are adopted in this paper when calculating the perturbations of the 2nd-degree earth tides on LAGEOS and LARES 2. The expressions under other conditions are listed in Tables 1 and 2 of Kaula (1966).

# 3. Frequency response of the satellite orbit dependent with the tidal deformation of the earth

The preceding derivation of the expression for the gravitational potential of Earth's tides (Equation 6) adopted an idealized frequency-independent assumption for the Love numbers . However, the tidal response of the actual Earth exhibits significant frequency dependence: anelasticity of the mantle leads to phase lag in tidal deformation, and this effect varies with the tidal frequency band (e.g., long-period, diurnal, semidiurnal) (Wahr and Bergen 1986; Petit and Luzum 2010); meanwhile, resonance effects from Earth's rotation, oblateness, and Free Core Nutation (FCN) significantly modulate the Love number values. Ignoring this frequency dependence would lead to modeling errors in the amplitudes of tidal perturbations for different frequency bands, which in turn would affect the extraction of the frame-dragging effect at sub-percent level accuracy.

To precisely quantify this characteristic, it is necessary to introduce frequency-dependent astronomical amplitudes $H_l^m(f)$ and Love numbers $k_{lm}(f)$. The astronomical amplitudes (Cartwright and Tayler 1971; Cartwright and Edden 1973) are obtained by analyzing ephemerides to determine the contribution of the tidal potential's equivalent tide height for different frequencies $f$. The frequency-dependent Love numbers $k_{lm}(f)$ are constructed by strictly adhering to the layered modeling framework of IERS Conventions 2010: the basic elastic parameters are derived from the elliptical, rotating, non-oceanic version of the Preliminary Reference Earth Model (PREM) (Dziewonski and Anderson 1981)— a foundational setup for body tide modeling first formalized by Wahr (1981), who explicitly incorporated Earth's ellipticity and rotation into tidal Love number calculations; the frequency response due to mantle anelasticity is quantified using the mantle Q model of Widmer et al. (1991) in conjunction with the power-law relationship of Wahr and Bergen (1986); resonance effects for diurnal tides are addressed using the nutation-wobble coupling model of Mathews et al. (2002), which incorporates corrections for the resonance frequencies ($\sigma_\alpha$) and coefficients ($L_\alpha$) of the Chandler wobble, Free Core Nutation (FCN), and Free Inner Core Nutation (FICN), as Equation 7 shows; and the frequency-dependent contribution of ocean tide loading is converted into corrections for the body tide Love numbers via the equivalent model of Wahr and Sasao (1981), with ocean tide data sourced from the FES2004 model (Lyard et al. 2006). The coupled application of these models ensures the computational accuracy of both the real part (amplitude) and the imaginary part (energy dissipation) of the Love numbers across different tidal frequency bands (long-period, diurnal, and semidiurnal).

$$\begin{cases} k_{21}^{(0)}(f) = k_{21}^{(0)} + \sum_{\alpha=1}^{3} \frac{L_\alpha}{\sigma - \sigma_\alpha} \\ \sigma = f/(15 \times 1.002737909) \end{cases} \tag{7}$$

Figure 2 is a comparative diagram of the frequency-dependent tidal perturbations (see in 2.1.5 and Appendix Table 1) and frequency-independent tidal perturbations (nominal love numbers for each degree and order are referred to Petit and Luzum (2010) on the node of LARES 2, and it can be observed from Figure 2(b) that the amplitude difference between the two is considerable, reaching up to 322 mas—which is more than ten times the frame-dragging effect—thus, it is essential to take into account the frequency-dependent response of the Earth tides when calculating satellite orbit perturbations.

Based on this, the term in Equation 6 related to the tide-generating body, $\frac{GM^*}{a*}\left(\frac{R}{a^*}\right)^l$ must be transformed into a standard form that includes frequency information, with the specific substitution as $\frac{GM^*}{a*}\left(\frac{R}{a^*}\right)^l = gN_{lm}H_l^m(f)$.

$g = \frac{GM}{R^2}$ is the acceleration of gravity at Earth's surface. $H_l^m(f)$ is proportional to $(M^*/M)(R/a^*)^l$, with dimensions of $L$, and it incorporates the inclination functions $F_{lmh}(i^*)$ and eccentricity functions $G_{lhj}(e^*)$ of the tide-generating body.

In tidal theory, the tidal potential at Earth's surface is proportional to the product of the surface gravity acceleration and the equivalent tide height, i.e., $gH_l^m(f)$, and the spherical harmonic normalization coefficient $N_{lm} = \sqrt{\dfrac{(2l+1)(l-m)!}{4\pi(l+m)!}}$ originates from spherical harmonic theory, without it, the amplitudes of different $l,m$ components cannot be directly compared.

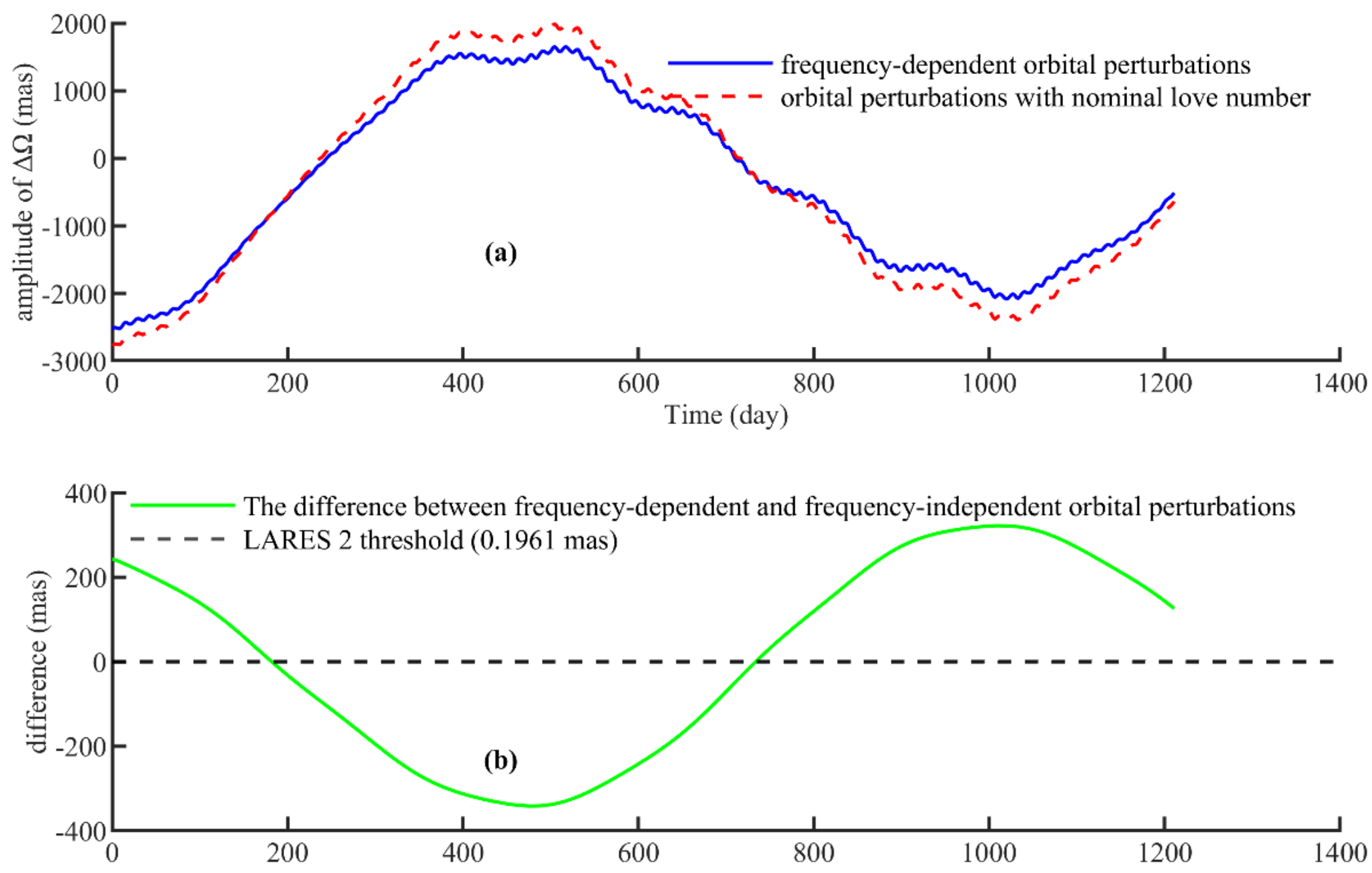

**Fig. 2** Comparison of tidal perturbation time series on node of LARES 2: frequency-dependent and nominal love numbers.

In summary, the additional gravitational potential due to Earth's tidal deformation, which perturbs an artificial satellite, is expressed as:

$$V_{\mathcal{T},\mathcal{S}} = -g\sum_{l=2}^{+\infty}\left(\frac{R}{a}\right)^{l+1}\sum_{m=0}^{l} N_{lm} \times$$
$$\sum_{p=0}^{l} F_{lmp}(i)\sum_{q=-\infty}^{+\infty} G_{lpq}(e)\sum_{f} H_l^m(f)k_{lm}(f)\cos\left(\sigma t + \psi_{lmpq} - \epsilon_{lm}(f)\right) \quad (8)$$

where $\sigma = 2\pi f$ is the angular frequency of the tidal component of the tide-generating body (satisfying $\sigma = 2\pi / P$ with period $P$), which is a combination angle variable composed of the orbital elements of the tide-generating body. To efficiently and standardly encode all frequency components, Doodson numbers are used as coefficients for the corresponding linear combination of the angular frequencies of the tide-generating body, with the expression given by Equation 9. $\epsilon_{lm}(f)$represents the phase lag of the Earth tide for degree order $l,m$ and frequency $\sigma$, caused by the anelasticity of the Earth's mantle, and can be treated as a time-invariant quantity in this study.

$$\sigma = j_1\dot{\tau} + j_2\dot{s} + j_3\dot{h} + j_4\dot{p} + j_5\dot{N}' + j_6\dot{p}_s \quad (9)$$

Here, $j_1 \sim j_6$ are the Doodson numbers (Doodson 1921) that define the tidal components; for example, $j_1$ is used to distinguish the types of tides: $j_1 = 1$ for long-period/zonal tides, $j_1 = 1$ for diurnal/tesseral tides, and $j_1 = 2$ for semidiurnal/sectorial tides. $\dot{\tau}$, $\dot{s}$, $\dot{h}$, $\dot{p}$, $\dot{N}'$ and $\dot{p}_s$ are respectively the rate of change of the lunar mean time, the Moon's mean longitude, the Sun's mean longitude, the Moon's mean longitude of perigee, the Moon's mean longitude of the ascending node, and the Sun's mean longitude of perigee.

These six fundamental astronomical angular frequencies can be expressed as $\alpha_i = \{347.80925061, 13.17639673, 0.98564734, 0.11140408, 0.05295392, 0.00004707\}$ ( $\circ/day$ ), with their corresponding periods being $\gamma_i = 360/\alpha_i$ (days). Thus, $\sigma = \sum_{i=1}^{6} 2\pi j_i/\gamma_i$.

For the time variation of the phase term $\sigma t + \psi_{lmpq} - \epsilon_{lm}(f)$, it should be noted that $\psi_{lmpq}$ is also a time-varying quantity, which can be written as $\left[m(\dot{\Omega}-\dot{\theta})+(l-2p)\dot{\omega}+(l-2p+q)\dot{\mu}\right]t$. Therefore, this phase can be expressed as $\left[\sigma+m(\dot{\Omega}-\dot{\theta})+(l-2p)\dot{\omega}+(l-2p+q)\dot{\mu}\right]t-\varepsilon_{lm}(f)=\dot{\Gamma}_{jlmp}t-\varepsilon_{lm}(f)$. Taking the first derivative with respect to time of the phase term yields the frequency of the tidal perturbation (Bertotti and Carpino 1989):

$$\dot{\Gamma}_{j,lmp} = \sum_{i=1}^{6} \frac{2\pi j_i}{\gamma_i} + m\left(\dot{\Omega} - \dot{\theta}\right) + (l-2p)\dot{\omega} + (l-2p+q)\dot{\mu} \tag{10}$$

This study focuses on large-amplitude long-period perturbations: they require modeling, while short-period ones (negligible amplitude) average to zero over a full satellite orbit. That is, only the case where $l$-2$p$+$q$=0 is considered.

Secondly, to more comprehensively consider tidal perturbations with amplitudes within the observable range, in addition to focusing on the "source" of the astronomical amplitude of each tide—that is, calculating only the case when $l$=2 we also need to examine the eccentricity function: for the case where q=0, $G_{lpq}(e) \approx 1$, and the perturbation amplitude is not excessively scaled, thus requiring key consideration; for the case where q≠0, $G_{lpq}(e) \approx 0$, and the perturbation amplitude also becomes negligible, therefore this study does not consider it. In summary, this study focuses on the mode where $(l,p,q)=(2,1,0)$. At this time, the tidal perturbation frequency simplifies to:

$$\dot{\Gamma}_{j,lmp} = \sum_{i=1}^{6} \frac{2\pi j_i}{\gamma_i} + m\left(\dot{\Omega} - \frac{2\pi}{\gamma_{1+2}}\right) = \sum_{i=1}^{6} \frac{2\pi j_i}{\gamma_i} + m\left(\dot{\Omega} - \frac{2\pi}{360/(\alpha_1+\alpha_2)}\right) \tag{11}$$

Notably, incorporating other planets' tidal effects when deriving Earth tide gravitational potential would boost precision but increase complexity. Given their weak effects and the sharp rise in computational cost for tidal potential expansion, these effects are temporarily excluded here.

# 4. Tidal perturbation of the orbit inclination and node

Tidal perturbations originate from conservative gravitation, and in Lagrangian perturbation theory, the disturbing function $\mathcal{R}$ is the potential energy function of conservative forces; the tidal potential, as the potential function of tidal gravitation, is physically consistent with $\mathcal{R}$, therefore in Lagrange's equations, the tidal potential (minus) can replace $\mathcal{R}$ to analyze the perturbations of tides on satellite orbits:

$$\frac{d\Omega}{dt} = \frac{1}{\sqrt{GMa(1-e^2)}\sin i}\frac{\partial \mathcal{R}}{\partial i} = -\frac{1}{\sqrt{GMa(1-e^2)}\sin i}\frac{\partial V_{\mathcal{T},\mathcal{S}}}{\partial i} \tag{12}$$

By integrating the instantaneous rate of change of the right ascension of the ascending node $d\Omega/dt$ over time, the cumulative perturbation $\Delta\Omega_{lmpq}$ can be obtained (Kaula 1966; Bertotti and Carpino 1989):

$$\Delta\Omega_{lmpq} = \int \frac{d\Omega}{dt} dt = \frac{1}{\sin i} \sqrt{\frac{GMR^{2l-2}}{a^{2l+3}(1-e^2)}} \frac{\partial F_{lmp}(i)}{\partial i} G_{lpq}(e) N_{lm} \frac{H_l^m(f) k_{lm}(f)}{\dot{\Gamma}_{jlmp}} \sin\left(\dot{\Gamma}_{jlmp} t - \epsilon_{lm}(f)\right) \quad (13)$$

It can be seen that the cumulative perturbation $\Delta\Omega_{lmpq}$ is a quasi-trigonometric function. To obtain its amplitude, the maximum value of 1 is taken for $\sin\left(\dot{\Gamma}_{jlmp} t - \epsilon_{lm}(f)\right)$, yielding the amplitude:

$$\Delta\Omega_{lmpq} = \frac{1}{\sin i} \sqrt{\frac{GMR^{2l-2}}{a^{2l+3}(1-e^2)}} \frac{\partial F_{lmp}(i)}{\partial i} G_{lpq}(e) N_{lm} \frac{H_l^m(f) k_{lm}(f)}{\dot{\Gamma}_{jlmp}} \quad (14)$$

Similarly, the perturbation period of Earth tides on the orbital inclination is consistent with that on the right ascension of the ascending node, and the derivation of the perturbation amplitude is also analogous, with the only difference being in the Lagrange equations used. The tidal perturbations on other orbital elements can be derived in the same manner, and will not be elaborated further in this paper.

$$\frac{di_{lmpq}}{dt} = \frac{1}{\sqrt{GMa(1-e^2)} \sin i} \left( \cos i \frac{\partial V_{T,S}}{\partial \omega} - \frac{\partial V_{T,S}}{\partial \Omega} \right) \quad (15)$$

Here, the derivation process is omitted and the result is given directly:

$$\Delta i_{lmpq} = \frac{[\cos i\,(l-2p) - m]}{\sin i} \sqrt{\frac{GMR^{2l-2}}{a^{2l+3}(1-e^2)}} F_{lmp}(i) G_{lpq}(e) N_{lm} \frac{H_l^m(f) k_{lm}(f)}{\dot{\Gamma}_{jlmp}} \quad (16)$$

The orbital elements of LARES 2 and LAGEOS satellites used in the calculations are shown in Table 1, including the mean orbital inclination, mean semi-major axis, mean eccentricity (Ciufolini et al. 2023), nodal precession period $P(\Omega)$ — with the $P(\Omega)$ P(Ω) of both satellites being 1050 days as suggested by Ciufolini. These mean orbital elements cover a 127-day observation period from July 13, 2022, to November 16, 2022, and derived from SLR observations processed by the GEODYN II and UTOPIA orbit estimators.

**Table 1** The orbital elements of LARES 2 and LAGEOS

| | Inclination (degrees) | mean semi-major axis (km) | mean eccentricity | P(Ω) (day) |
|---|---|---|---|---|
| LARES 2 | 70.1615 | 12266.1359395 | 0.00027 | -1050 |
| LAGEOS | 109.8469 | 12270.020705 | 0.00403 | 1050 |

As for the calculation of the instantaneous phase of tidal perturbations, it is actually simpler. The essence is the superposition result of the dynamic accumulation of astronomical angles of tide-generating bodies and satellite orbital arguments with the phase lag of Earth's anelastic response. Its core physical significance is: at the target epoch, the tidal disturbance signal of tide-generating bodies on Earth, after being modulated by Earth's anelasticity, acts on the instantaneous phase state of the satellite orbit, that is, "absolute astronomical benchmark (such as J2000.0 epoch) + time evolution part + satellite relative orbital benchmark (first observation moment) + time evolution part + geophysical response (anelastic lag)", to ensure high precision and reproducibility of the phase. Among them, the Earth lag response part is the arctangent of the ratio of the imaginary part to the real part of the Love number.

# 5. Results of tidal perturbations of the orbits of LARES 2 and LAGEOS

### 5.1 Tidal perturbations exceeding the thresholds

For the 402 perturbations of earth tides (392 2-nd degree tides and 10 3-rd degree tides) to the orbits of LARES

2 and LAGEOS, their perturbation periods and amplitudes are detailed in Appendix Table 1. To identify significant tidal perturbation modes from these, it is necessary to set a threshold based on the orbit determination accuracy of the two satellites. Specifically, the threshold should refer to the intrinsic property of satellite orbit determination — i.e., the minimum resolvable capability of orbit solutions achievable with existing observational data and orbit determination methods (corresponding to internal consistency accuracy), rather than the observed fitting residuals containing biases from various physical perturbation models (corresponding to external consistency accuracy). The observed fitting residuals reflect the actual effect of implementing the model into orbit solutions; this indicator is used to evaluate the deviation adaptability between various physical perturbation models and actual observations, which is thus not suitable as a criterion for screening tidal perturbation modes. Based on this, this study adopts the RMS of overlap orbit differences for LAGEOS and LARES 2 (Shao et al. 2025).

For LAGEOS, the RMS of overlap orbit differences in the tangential (T) $\Delta r_t$ and normal (N) directions $\Delta r_n$ are 2.16 cm and 1.34 cm, respectively, from which the corresponding uncertainties in the orbital inclination ( $\Delta i_{un}$ ) and the node ( $\Delta\Omega_{un}$ ) are derived as 0.2253 mas and 0.2451 mas via Equation 17. For LARES 2, the mean RMS values of overlap orbit differences in the T and N directions are smaller (1.35 cm for $\Delta r_t$ and 1.14 cm for $\Delta r_n$, respectively) due to its favorable orbital and structural characteristics; correspondingly, $\Delta i_{un}$ and $\Delta\Omega_{un}$ are 0.1917 mas and 0.1961 mas, respectively. A tidal perturbation mode was retained if either its $\Delta i_{un}$ or $\Delta\Omega_{un}$ amplitude exceeded the threshold for the corresponding satellite. These values were adopted as a threshold to identify 83 significant tidal perturbation modes (81 2-nd degree tides and 2 3-rd degree tides). The perturbation periods and amplitudes for these selected modes are presented in Tables 2~4.

$$
\begin{cases}
\Delta\Omega_n = \dfrac{1}{a}\Delta r_n \sin i \\
\Delta\Omega_t = \dfrac{1}{a}\Delta r_t \cos i \\
\Delta\Omega_{un} = \dfrac{1}{a}\sqrt{{\Delta\Omega_n}^2 + {\Delta\Omega_t}^2} \\
\Delta i_{un} = \dfrac{\Delta r_n}{a}
\end{cases}
\tag{17}
$$

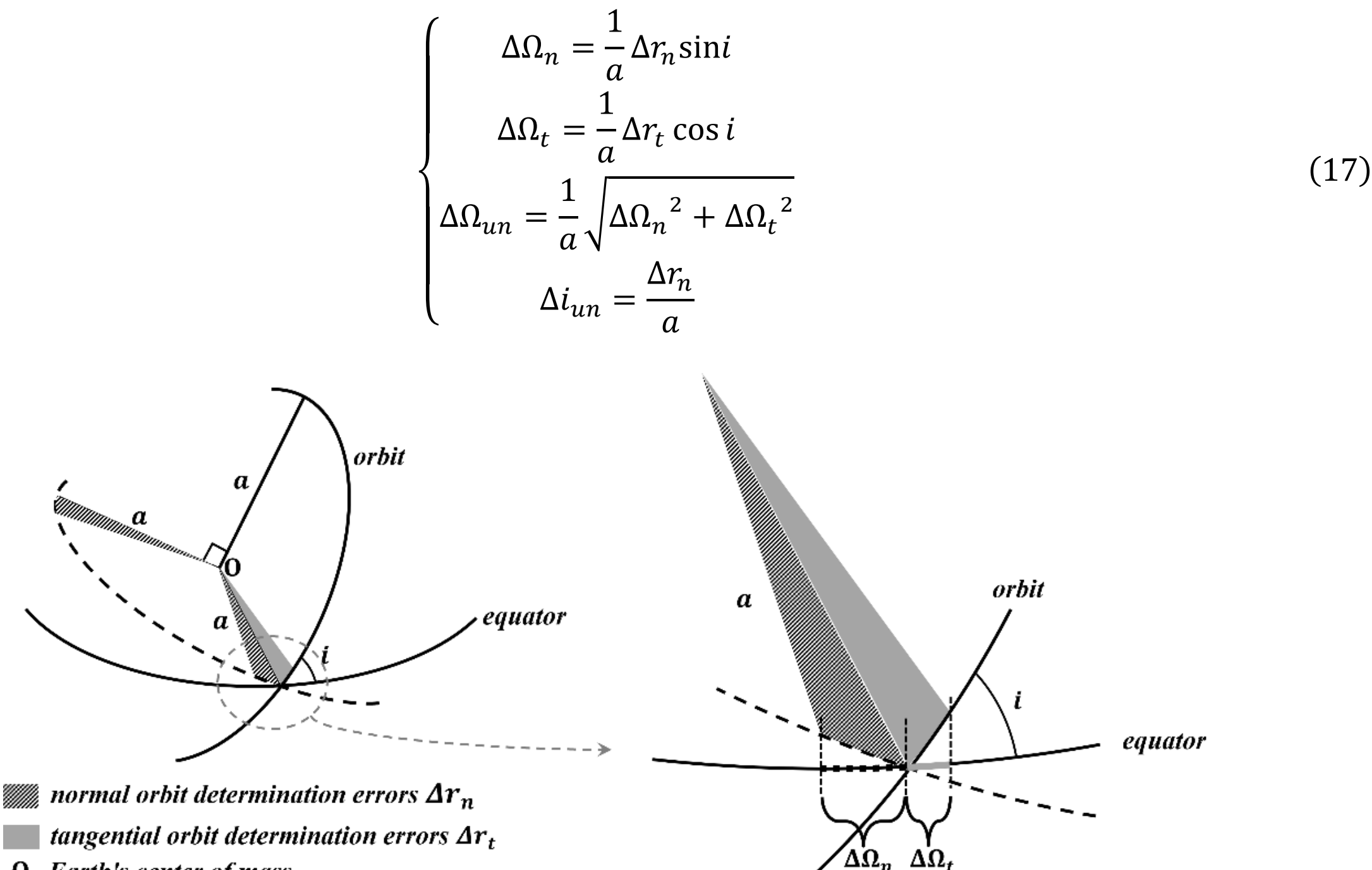

**Fig. 3** Schematic diagram of the contribution of satellite orbital normal and tangential errors to the node and inclination

**Table 2** Perturbation periods and amplitudes of the right ascension of the ascending node and orbital inclination for LAGEOS and LARES 2 satellites (l=2, m=0, p=1, q=0)

| Tide name | Doodson number | Love number | $H_l^m$(m) | Perturbation periods (days) | | $\Delta\Omega_{lmpq}$ (mas) | | $\Delta i_{lmpq}$ (mas) | |
|---|---|---|---|---|---|---|---|---|---|
| | | | | LAGEOS | LARES 2 | LAGEOS | LARES 2 | LAGEOS | LARES 2 |
| | 055.565 | 0.315416 | 0.02793 | 6798.3636 | 6798.3636 | -1073.8847 | 1074.6041 | 0 | 0 |
| | 055.575 | 0.313178 | -0.00027 | 3399.1818 | 3399.1818 | 5.1538 | -5.1573 | 0 | 0 |
| $S_a$ | 056.554 | 0.307390 | -0.00492 | 365.2596 | 365.2596 | 9.9050 | -9.9117 | 0 | 0 |
| $S_{sa}$ | 057.555 | 0.305946 | -0.031 | 182.6211 | 182.6211 | 31.0568 | -31.0776 | 0 | 0 |
| | 057.565 | 0.305896 | 0.00077 | 177.8438 | 177.8438 | -0.7511 | 0.7516 | 0 | 0 |
| | 058.554 | 0.305174 | -0.00181 | 121.7493 | 121.7493 | 1.2058 | -1.2067 | 0 | 0 |
| $M_{sm}$ | 063.655 | 0.302920 | -0.00673 | 31.8119 | 31.8119 | 1.1629 | -1.1637 | 0 | 0 |
| | 065.445 | 0.302709 | 0.00231 | 27.6667 | 27.6667 | -0.3469 | 0.3471 | 0 | 0 |
| $M_m$ | 065.455 | 0.302709 | -0.03518 | 27.5546 | 27.5546 | 5.2615 | -5.2651 | 0 | 0 |
| | 065.465 | 0.302699 | 0.00229 | 27.4433 | 27.4433 | -0.3411 | 0.3413 | 0 | 0 |
| | 065.655 | 0.302679 | 0.00188 | 27.0925 | 27.0925 | -0.2764 | 0.2766 | 0 | 0 |
| $M_{sf}$ | 073.555 | 0.301818 | -0.00583 | 14.7653 | 14.7653 | 0.4659 | -0.4662 | 0 | 0 |
| | 075.355 | 0.301728 | -0.00288 | 13.7773 | 13.7773 | 0.2147 | -0.2148 | 0 | 0 |
| $M_f$ | 075.555 | 0.301718 | -0.06663 | 13.6608 | 13.6608 | 4.9243 | -4.9276 | 0 | 0 |
| | 075.565 | 0.301718 | -0.02762 | 13.6334 | 13.6334 | 2.0372 | -2.0385 | 0 | 0 |
| $M_{tm}$ | 085.455 | 0.301197 | -0.01276 | 9.1329 | 9.1329 | 0.6294 | -0.6298 | 0 | 0 |
| | 085.465 | 0.301197 | -0.00529 | 9.1207 | 9.1207 | 0.2606 | -0.2607 | 0 | 0 |

The periods of zonal tides are determined solely by the orbital motion of the Sun and Moon, and are identical for both LAGEOS and LARES 2 satellites, with calculation results shown in Table 2. The typical characteristics of these tidal periods are: an 18.6-year (6798.3636 days) period associated with the regression of the Moon's nodes, and a half-year (182.6211 days) period associated with the variation in the Sun's declination. Their values depend on the long-term orbital parameters of the Sun and Moon, (such as the lunar nodal regression period and the annual solar motion).

For zonal tides, the perturbation amplitudes on the ascending nodes of LAGEOS and LARES 2 are nearly equal in magnitude but opposite in sign, whereas their perturbation amplitudes on orbital inclination are zero. This results from the axial symmetry of Earth's deformation induced by zonal tides: at a given latitude, solid-Earth uplift or subsidence and ocean mass redistribution are identical for all longitudes and thus do not depend on the right ascension of the ascending node. Consquently, in Equation 15, $\partial V_{T,S}/\partial\Omega = 0$, and with l-2p=0, the result of integrating $di_{lmpq}/dt$ over time is zero. This agrees with the conclusion of Bertotti and Carpino (1989) that "zonal tides (m=0) cancel out due to symmetry".

However, due to differences in data recency, the numerical results in this study differs from those in the 1989 report by approximately 1%-20%. At that time, LARES 2 had not yet been launched, and its orbital parameters were simulated based on LAGEOS, whereas today several years of laser-ranging observations are available for LARES 2, directly affecting amplitude estimates (for example, the semi-major axis differs by about 4 km). In addition, the Love numbers used in 1989 adopted a uniform value of $k_2$=0.317, ignoring frequency dependence. In contrast, the present work employs frequency-dependent Love numbers constrained by observational studies of Earth's internal structure (e.g., mantle viscosity), providing a more realistic representation of Earth's tidal response. Similar differences apply to the cases following.

Furthermore, as shown in Table 2, the $\Omega_1$ tide (055.565) exhibits the largest perturbation amplitude on the ascending nodes of the two satellites. If we set aside the condition that the symmetric configuration of LARES 2 and LAGEOS cancels out the effect of the $\Omega_1$ tide for the scientific mission of verifying the frame-dragging effect (following the same principle as the cancellation of $J_2$'s influence), the accurate modeling of $\Omega_1$ tidal perturbations would be crucial for the precise orbit determination of each individual satellite. To illustrate this, if we use the Love number of the $\Omega_1$ tide inverted by Minkang Cheng (2025) through long-term observational data from multiple SLR satellites—calculated as $\sqrt{(0.3022+0.01375)^2+(-0.00553)^2}=0.315998$ — to compute the perturbation amplitude of this tide on the nodes of the two satellites, we obtain -1075.8675 mas (for LAGEOS) and 1076.5883 mas(for LARES 2), respectively. These values differ from the results derived using the IERS 2010 recommended model (with $k_{18.6yr}$=0.30190+0.01347-i0.00541) by -1.9828 mas and 1.9842 mas, which are significantly larger than their respective thresholds of 0.2451 mas and 0.1961 mas—indicating the differences are statistically observable (otherwise, the inverted values would be deemed indistinguishable from the model values). For maintaining millimeter-level orbit determination accuracy and high-precision missions lacking symmetric satellite configurations to cancel $J_2$ and zonal tides effects, without proper correction, these discrepancies will accumulate significantly over time due to orbit integration, manifesting as systematic biases. These differences not only highlight the limitations of the IERS 2010 model in capturing frequency-dependent tidal responses and deep Earth dynamic effects (e.g., core-mantle coupling) but also underscore the necessity of updating tidal parameters with long-term, high-precision observational data.

**Table 3** Perturbation periods and amplitudes of the right ascension of the ascending node and orbital inclination for LAGEOS and LARES 2 satellites (l=2, m=1, p=1, q=0)

| Tide name | Doodson number | Love number | $H_l^m$(m) | Perturbation periods (days) | | $\Delta\Omega_{lmpq}$(mas) | | $\Delta i_{lmpq}$(mas) | |
|---|---|---|---|---|---|---|---|---|---|
| | | | | LAGEOS | LARES 2 | LAGEOS | LARES 2 | LAGEOS | LARES 2 |
| $2Q_1$ | 125.755 | 0.298013 | -0.00664 | -6.9045 | -6.8149 | 0.2410 | 0.2382 | -0.1000 | 0.0988 |
| $\sigma_1$ | 127.555 | 0.298003 | -0.00802 | -7.1441 | -7.0482 | 0.3012 | 0.2975 | -0.1250 | 0.1234 |
| | 135.645 | 0.297853 | -0.00947 | -9.2006 | -9.0421 | 0.4577 | 0.4504 | -0.1900 | 0.1868 |
| $Q_1$ | 135.655 | 0.297843 | -0.05020 | -9.2131 | -9.0542 | 2.4297 | 2.3908 | -1.0084 | 0.9916 |
| $\rho_1$ | 137.455 | 0.297813 | -0.00954 | -9.6446 | -9.4707 | 0.4833 | 0.4752 | -0.2006 | 0.1971 |
| | 145.545 | 0.297483 | -0.04946 | -13.8127 | -13.4586 | 3.5847 | 3.4972 | -1.4877 | 1.4505 |
| $O_1$ | 145.555 | 0.297473 | -0.26221 | -13.8409 | -13.4853 | 19.0422 | 18.5766 | -7.9028 | 7.7050 |
| $\tau_1$ | 147.555 | 0.297393 | 0.00343 | -14.9759 | -14.5605 | -0.2694 | -0.2623 | 0.1118 | -0.1088 |
| $N_{\tau 1}$ | 153.655 | 0.296623 | 0.00194 | -24.5007 | -23.4083 | -0.2487 | -0.2379 | 0.1032 | -0.0987 |
| $L_{k1}$ | 155.455 | 0.296363 | 0.00741 | -27.8101 | -26.4111 | -1.0772 | -1.0243 | 0.4471 | -0.4249 |
| $N_{O1}$ | 155.655 | 0.296333 | 0.02062 | -28.2971 | -26.8499 | -3.0498 | -2.8975 | 1.2657 | -1.2018 |
| | 155.665 | 0.296323 | 0.00414 | -28.4154 | -26.9564 | -0.6149 | -0.5840 | 0.2552 | -0.2422 |
| $\chi_1$ | 157.455 | 0.295993 | 0.00394 | -32.8059 | -30.8765 | -0.6748 | -0.6359 | 0.2801 | -0.2638 |
| $\pi_1$ | 162.556 | 0.289961 | -0.00714 | -137.7180 | -109.0991 | 5.0290 | 3.9890 | -2.0871 | 1.6545 |
| | 163.545 | 0.287131 | 0.00137 | -214.1084 | -152.0845 | -1.4856 | -1.0566 | 0.6165 | -0.4382 |
| $P_1$ | 163.555 | 0.286921 | -0.12203 | -221.0708 | -155.5646 | 136.5268 | 96.1938 | -56.6609 | 39.8982 |
| | 164.554 | 0.280660 | 0.00103 | -560.0173 | -270.9718 | -2.8555 | -1.3834 | 1.1851 | -0.5738 |
| $S_1$ | 164.556 | 0.280660 | 0.00289 | -560.0993 | -270.9910 | -8.0131 | -3.8819 | 3.3256 | -1.6101 |
| | 165.345 | 0.267821 | 0.00007 | 5365.5215 | -581.9412 | 1.7742 | -0.1927 | -0.7363 | -0.0799 |
| | 165.535 | 0.262000 | 0.00005 | 1519.3123 | -802.2016 | 0.3511 | -0.1856 | -0.1457 | -0.0770 |
| | 165.545 | 0.259851 | -0.00730 | 1241.7937 | -909.5249 | -41.5484 | 30.4698 | 17.2433 | 12.6379 |
| $K_1$ | 165.555 | 0.257463 | 0.36878 | 1050.0000 | -1050.0000 | 1758.4459 | -1760.6741 | -729.7842 | -730.2731 |
| | 165.565 | 0.254755 | 0.05001 | 909.5249 | -1241.7937 | 204.3862 | -279.4065 | -84.8237 | -115.8892 |
| | 165.575 | 0.251757 | -0.00108 | 802.2016 | -1519.3123 | -3.8472 | 7.2956 | 1.5967 | 3.0260 |
| $\psi_1$ | 166.554 | 0.526244 | 0.00293 | 270.9910 | 560.0993 | 7.3700 | 15.2520 | -3.0587 | 6.3261 |
| | 166.556 | 0.526104 | -0.00004 | 270.9718 | 560.0173 | -0.1056 | -0.2185 | 0.0438 | -0.0906 |
| | 166.564 | 0.466726 | 0.00005 | 260.6030 | 517.4666 | 0.1073 | 0.2133 | -0.0445 | 0.0885 |
| | 167.355 | 0.335867 | 0.00018 | 172.1381 | 256.1130 | 0.1836 | 0.2735 | -0.0762 | 0.1134 |
| $\phi_1$ | 167.555 | 0.328564 | 0.00525 | 155.5646 | 221.0708 | 4.7331 | 6.7347 | -1.9643 | 2.7933 |
| | 167.565 | 0.327234 | -0.00020 | 152.0845 | 214.1084 | -0.1756 | -0.2475 | 0.0729 | -0.1026 |
| | 168.554 | 0.314689 | 0.00031 | 109.0991 | 137.7180 | 0.1877 | 0.2373 | -0.0779 | 0.0984 |
| $\theta_1$ | 173.655 | 0.302004 | 0.00395 | 30.8765 | 32.8059 | 0.6497 | 0.6911 | -0.2696 | 0.2867 |
| $J_1$ | 175.455 | 0.301544 | 0.02062 | 26.8499 | 28.2971 | 2.9447 | 3.1073 | -1.2221 | 1.2888 |
| | 175.465 | 0.301534 | 0.00409 | 26.7443 | 28.1798 | 0.5818 | 0.6138 | -0.2414 | 0.2546 |
| $S_{O1}$ | 183.555 | 0.300244 | 0.00342 | 14.5605 | 14.9759 | 0.2637 | 0.2716 | -0.1094 | 0.1126 |
| $O_{O1}$ | 185.555 | 0.300144 | 0.01129 | 13.4853 | 13.8409 | 0.8060 | 0.8283 | -0.3345 | 0.3436 |
| | 185.565 | 0.300144 | 0.00723 | 13.4586 | 13.8127 | 0.5151 | 0.5294 | -0.2138 | 0.2196 |

The periods of tesseral tides are coupled with the nodal period Ω of the satellite. Certain tidal components—most notably the $K_1$ tide—have period that coincide exactly with the nodal precession period, potentially interfering with the relativistic precession measurements. In calculating these periods, the direction of motion of the tidal component must also be considered (the sign before the period indicates its direction). The computed values are presented in Table 3. For example, the perturbation period of the $K_1$ tide on the orbits of both LARES 2 and LAGEOS coincides with the nodal precession periods of the two satellites (1050 days).

A comparison of the perturbation periods for both satellites in each row of Table 3 that when m≠0, the frequency coupling term $m(\dot{\Omega} - \dot{\theta})$ involving the satellite's nodal precession and Earth's rotation begins to contribute significantly to the inter-satellite asymmetry in tidal perturbation frequencies. Consequently, differences arise between the prograde and retrograde orbits. As a result, the effects of tesseral tides on Ω and $i$ cannot be eliminated through a simple linear combination of two satellites. Instead, separate and precise tidal perturbation models must be developed for each satellite individually.

**Table 4.** Perturbation periods and amplitudes of the node and inclination for LAGEOS and LARES 2 satellites (l=2, m=2, p=1, q=0)

| Tide name | Doodson number | Love number | $H_l^m$(m) | Perturbation periods (days) | | $\Delta\Omega_{lmpq}$ (mas) | | $\Delta i_{lmpq}$ (mas) | |
|---|---|---|---|---|---|---|---|---|---|
| | | | | LAGEOS | LARES 2 | LAGEOS | LARES 2 | LAGEOS | LARES 2 |
| | 235.755 | 0.301083 | 0.01601 | -6.9502 | -6.7709 | 0.2452 | -0.2391 | 0.6794 | 0.6627 |
| | 237.555 | 0.301083 | 0.01932 | -7.1930 | -7.0012 | 0.3063 | -0.2983 | 0.8486 | 0.8269 |
| | 245.555 | 0.301083 | -0.00389 | -9.2680 | -8.9519 | -0.0795 | 0.0768 | -0.2201 | -0.2129 |
| | 245.645 | 0.301083 | -0.00451 | -9.2819 | -8.9649 | -0.0923 | 0.0892 | -0.2556 | -0.2472 |
| $N_2$ | 245.655 | 0.301083 | 0.12099 | -9.2946 | -8.9768 | 2.4785 | -2.3954 | 6.8667 | 6.6394 |
| | 247.455 | 0.301063 | 0.02298 | -9.7340 | -9.3860 | 0.4930 | -0.4757 | 1.3658 | 1.3184 |
| | 255.545 | 0.301063 | -0.02358 | -13.9969 | -13.2883 | -0.7274 | 0.6910 | -2.0152 | -1.9153 |
| $M_2$ | 255.555 | 0.301063 | 0.63192 | -14.0257 | -13.3143 | 19.5330 | -18.5547 | 54.1159 | 51.4291 |
| | 263.655 | 0.301063 | -0.00466 | -25.0861 | -22.8978 | -0.2576 | 0.2353 | -0.7138 | -0.6522 |
| | 265.455 | 0.301063 | -0.01786 | -28.5667 | -25.7630 | -1.1244 | 1.0147 | -3.1151 | -2.8126 |
| | 265.555 | 0.301063 | 0.00359 | -28.8215 | -25.9701 | 0.2280 | -0.2056 | 0.6318 | 0.5699 |
| | 265.655 | 0.301063 | 0.00447 | -29.0809 | -26.1805 | 0.2865 | -0.2581 | 0.7937 | 0.7153 |
| | 265.665 | 0.301063 | 0.00197 | -29.2058 | -26.2817 | 0.1268 | -0.1142 | 0.3513 | 0.3165 |
| | 271.557 | 0.301063 | 0.00070 | -110.5386 | -77.7839 | 0.1705 | -0.1201 | 0.4724 | 0.3328 |
| $T_2$ | 272.556 | 0.301063 | 0.01720 | -158.5079 | -98.8303 | 6.0084 | -3.7488 | 16.6462 | 10.3907 |
| $S_2$ | 273.555 | 0.301063 | 0.29400 | -280.0292 | -135.4907 | 181.4391 | -87.8472 | 502.6752 | 243.4916 |
| $R_2$ | 274.554 | 0.301063 | -0.00246 | -1200.0795 | -215.3872 | -6.5062 | 1.1685 | -18.0253 | -3.2388 |
| | 274.556 | 0.301063 | 0.00062 | -1200.4562 | -215.3993 | 1.6403 | -0.2945 | 4.5444 | 0.8163 |
| | 274.566 | 0.301063 | -0.00004 | -1457.8909 | -222.4473 | -0.1285 | 0.0196 | -0.3561 | -0.0544 |
| | 275.455 | 0.301063 | 0.00019 | 626.8389 | -451.6268 | -0.2625 | -0.1892 | -0.7272 | 0.5245 |
| | 275.545 | 0.301063 | 0.00103 | 568.9358 | -487.3636 | -1.2915 | -1.1070 | -3.5780 | 3.0684 |
| $K_2$ | 275.555 | 0.301063 | 0.07996 | 525.0000 | -525.0000 | -92.5151 | -92.5770 | -256.3120 | 256.6015 |
| | 275.565 | 0.301063 | 0.02383 | 487.3636 | -568.9358 | -25.5951 | -29.8991 | -70.9111 | 82.8733 |
| | 277.555 | 0.301063 | 0.00063 | 135.4907 | 280.0292 | -0.1881 | 0.3891 | -0.5212 | -1.0784 |
| | 275.575 | 0.301063 | 0.00259 | 454.7624 | -620.8969 | -2.5958 | -3.5464 | -7.1915 | 9.8298 |
| | 285.455 | 0.301063 | 0.00447 | 26.1805 | 29.0809 | -0.2579 | 0.2867 | -0.7145 | -0.7946 |
| | 285.465 | 0.301063 | 0.00195 | 26.0800 | 28.9570 | -0.1121 | 0.1245 | -0.3105 | -0.3452 |

For sectorial earth tides, the computed results are presented in Table 4. Among all sectorial tides, the $K_2$ and $S_2$ components are the most critical: they exhibit both the largest perturbation amplitudes and the longest periods. This means that any inaccuracies in the modeling these two tides (i.e., an inability to precisely represent their effects on satellite orbits) can significantly compromise the high-precision measurement of relativistic effects on the satellites' orbits.

In addition, for the perturbations of the 3rd-degree Earth tides, we calculated the 10 significant tidal constituents with astronomical amplitudes of no less than 0.001 m listed in Tables 5(a) to 5(d) of Cartwright and Tayler (1971), whose Doodson numbers are 065.555, 135.555, 155.555, 175.555, 235.655, 245.555, 265.555,

345.655, 355.555 and 375.555, respectively. The mode with $(l,p,q)=\{(3,0,0),(3,1,0),(3,2,0),(3,3,0)\}$ were adopted for the calculations. For the perturbations of the ascending node, only the two modes to LARES 2 reach the magnitude of tenths of a milliarcsecond: 265.555 for (l,m,p,q)=(3,2,1,0) and 065.555 for the (l,m,p,q)=(3,2,2,0), with the values being -0.3285 mas (-69.9868 days) and 0.3831 mas (80.7517 days), respectively. For the perturbations of the orbital inclination, none of the above tidal constituents attains the preset threshold. The perturbation amplitudes of the remaining 3rd-order Earth tide modes are extremely small, falling in the range of $10^{-10}$ to $10^{-5}$ mas, which are far below the screening threshold; therefore, no further elaboration on these modes is presented in this paper.

### 5.2 Limitations of tidal constituent screening methods and improvement strategies

The screening method using the threshold calculated from the RMS of overlap orbit differences, which objectively measures the contribution of individual tidal constituents, has inherent shortcomings — the cumulative contribution of the filtered secondary tidal constituents to the target orbital elements exceeds the aforementioned threshold, as shown in Figs. 4 and 5. This reflects the general drawbacks of such individual tidal constituent screening approaches: the coherent temporal superposition of massive tiny constituents (In particular, secondary tidal perturbations with perturbation periods of half a month and about a month), frequency cluster-orbital resonance coupling, and collective detectability may lead to the total superimposed effect exceeding the threshold, a phenomenon particularly significant for high-precision satellites like LAGEOS/LARES 2. Therefore, it is necessary to reconsider incorporating these secondary tidal constituents into the tidal perturbation modeling framework and evolve the selection principle from "isolated constituent judgment" to "collective superimposed effect evaluation". Specifically, based on the internal consistency accuracy of orbit determination (RMS of overlap orbit differences), the threshold should be redefined as the "minimum negligible limit of total collective effect", extending to dual-dimensional screening (time domain + frequency domain). In practice, a closed-loop strategy should be adopted: hierarchical coarse screening by Doodson-coded frequency clusters, verification of total superimposed effect, a dual-threshold system (single-constituent amplitude + total superimposed effect), and orbit determination back-substitution validation. The core criterion is that only when the total superimposed effect of constituent clusters is lower than the orbit determination accuracy threshold can the relevant constituents be safely excluded, which is consistent with the IERS 2010 Conventions and the requirements of high-precision satellite orbit determination.

Naturally, minimizing the difference between the satellite ascending node time series (after deducting various classical perturbations) and the theoretical frame dragging effect is only the original intention of this work. When evaluating the final results, there are two types of assessment methods: one based on fitting residuals (a more direct method starting from numerical results) and the other based on the uncertainty of physical models (a more rigorous method that requires propagating the uncertainty of each tidal constituent to the final orbital residuals via the law of error propagation). Thus, from the perspective of the error propagation assessment method, it is not better to deduct more tidal constituents, but to achieve a balance that considers both numerical performance and uncertainty—i.e., maximizing numerical gain while minimizing uncertainty propagation. However, strictly evaluating the uncertainty of each tidal constituent is extremely challenging; a feasible compromise is to use statistical resampling methods such as Monte Carlo simulation for estimation.

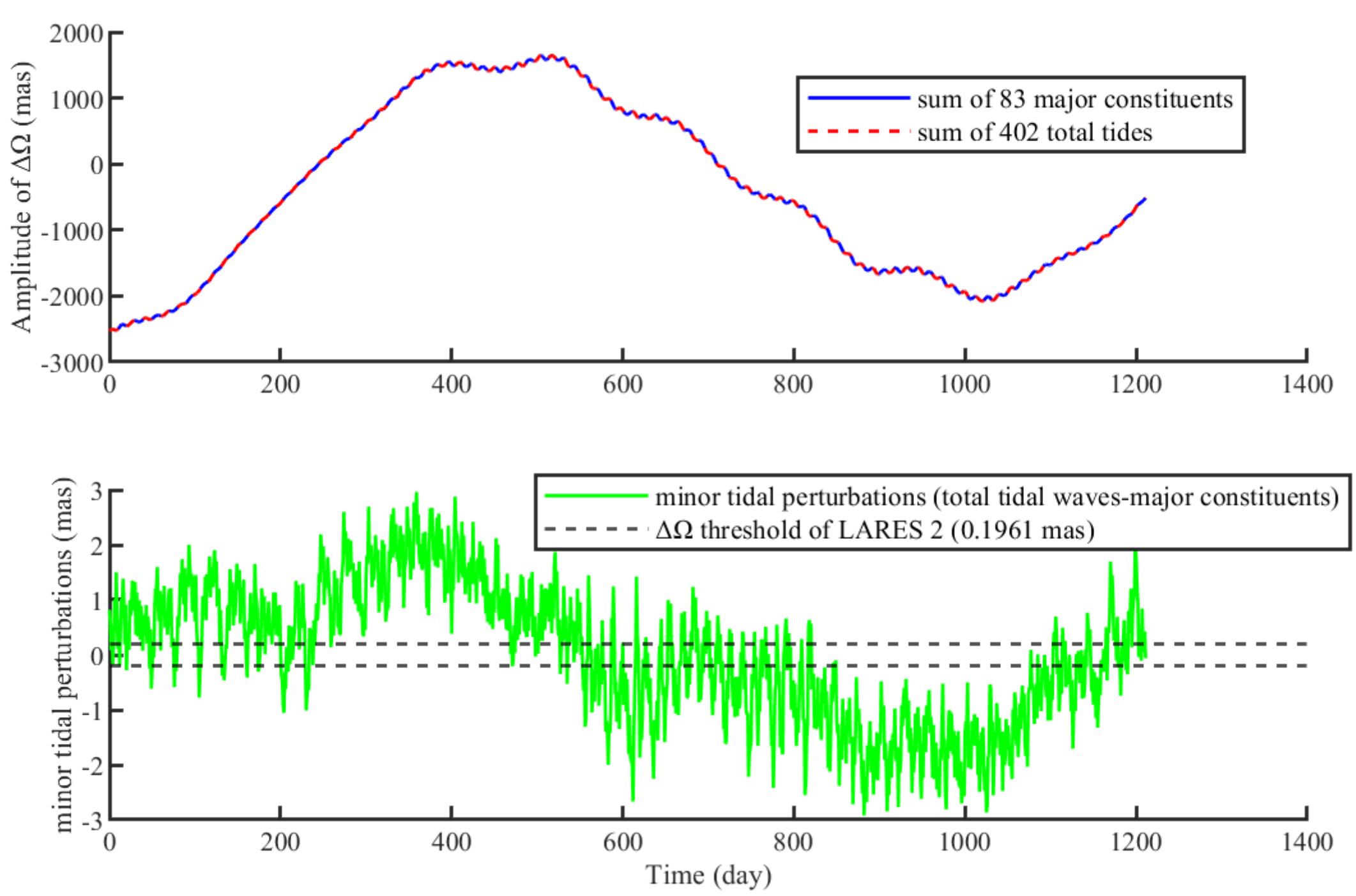

**Fig. 4** Comparison of tidal perturbations on Ω of LARES 2: sum of 83 major constituents vs. sum of 402 total tides

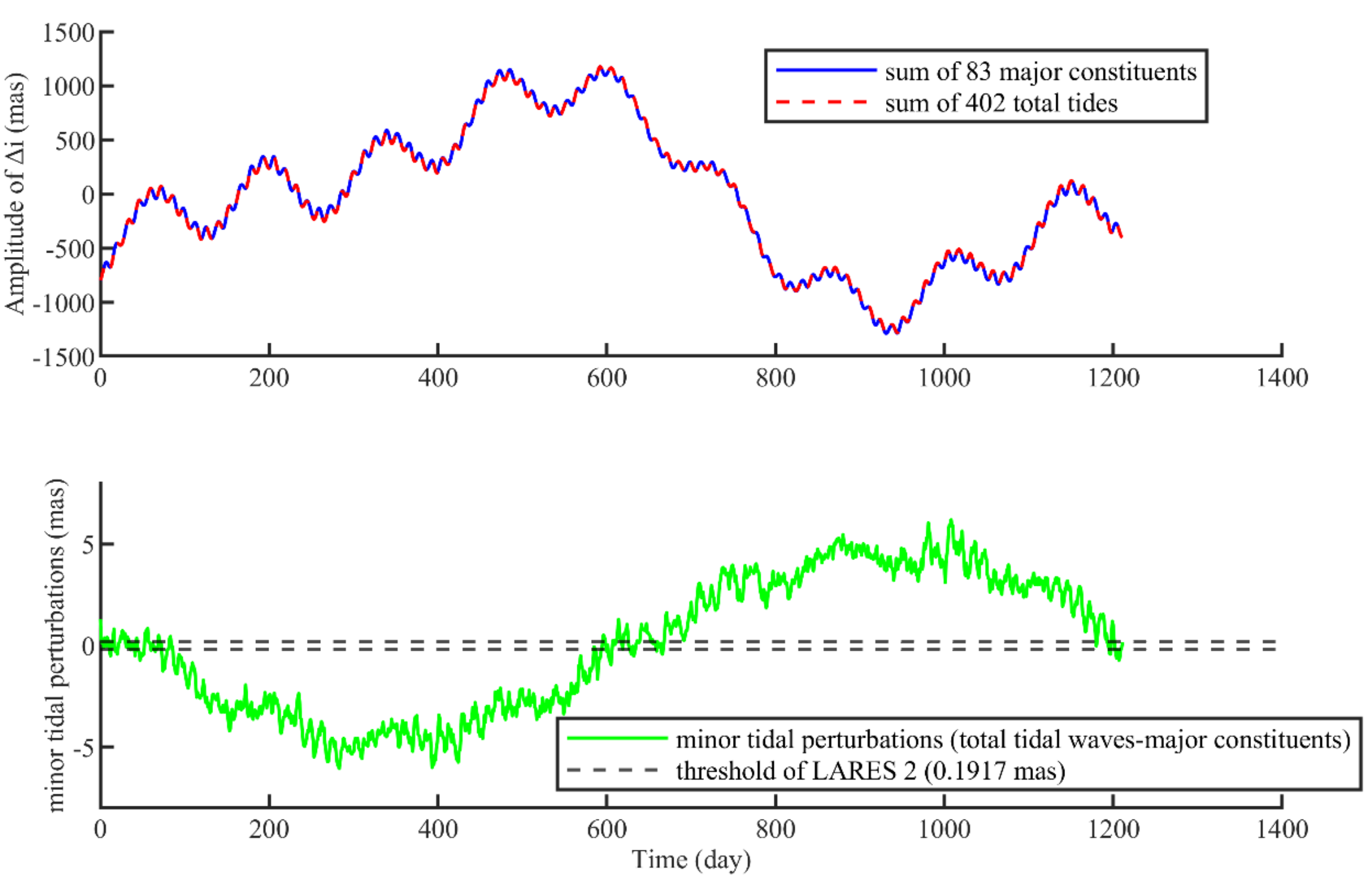

**Fig. 5** Comparison of tidal perturbations on i of LARES 2: sum of 83 major constituents vs. sum of 402 total tides

In addition to the screening approach, the impact of tidal perturbations on the accuracy of the Lense-Thirring effect detection arises from two sources: (1) limitations in the tidal perturbation model itself (for example, discrepancies between Kaula's linear perturbation theory and the actual tidal perturbations), and (2) uncertainties inherent in the tides (such as errors in Love numbers or in ocean tide models). The assessment of how to correct for tidal perturbations affects the detection accuracy of the Lense-Thirring effect is an important direction for future work and lies beyond the scope of this paper.

# 6. Conclusions

This study, based on Kaula perturbation theory and Lagrange's equations, systematically derived the formulas for Earth tidal perturbations on the right ascension of the ascending node ( $\Omega$ ) and orbital inclination ( $i$ ) of LARES 2 and LAGEOS satellites. By selecting 83 earth tides with significant tidal perturbation modes from 402 earth tide constituents, it quantitatively analyzed the asymmetric tidal perturbation characteristics between the two satellites, providing critical support for high-precision General Relativity verification and satellite orbital dynamics research.

Essentially, tidal perturbation asymmetry arises from differing relative angular velocities between satellite nodal precession and Earth's rotation for prograde/retrograde orbits; consequently, the same tide exhibits period and amplitude asymmetry on two satellites due to these frequency differences, disrupting the classical perturbation cancellation effect of the "butterfly configuration" and requiring individual modeling of $m \geq 1$ tesseral/sectorial tides (e.g., $K_1$ and $S_2$ ) for each satellite.

Further, differential response characteristics by tidal type: Zonal tides exhibit consistent periods and "equal in magnitude but opposite in sign" $\Delta\Omega$ responses, allowing partial cancellation through combination. Tesseral and Sectorial tides show period and amplitude difference. Hierarchy of perturbation source contributions: Earth tides are dominated by l=2 order components, with $K_1$ , $S_2$ as the primary interference sources. However, the cumulative contribution of the filtered secondary tidal constituents to the target orbital elements should not be ignored.

Building on these conclusions, future research will focus on three directions. First, this study is based on linear tidal theory, neglecting nonlinear interactions of tides. Future work should derive nonlinear tidal perturbation potentials through fluid dynamics equations, introducing nonlinear tidal coefficients to quantify nonlinear effects on satellite orbits. Second, the conventional constant-amplitude threshold for Earth tide componets selection is limited, tiny constituents'coherent superposition and frequency-orbital resonance push total effect over the threshold, so we propose shifting to collective effect evaluation via a time-frequency dual-dimensional closed-loop strategy, excluding only clusters with total effect below orbit determination accuracy. Specifically, implement hierarchical frequency cluster screening, dual-threshold verification and orbit back-substitution validation. Additionally, Long-term observational validation: LARES 2 has accumulated approximately 3 years of laser ranging data since its launch in July 2022. Future work should combine this with LAGEOS's long-term data to invert actual tidal perturbation amplitudes and periods through orbital fitting, validating theoretical model accuracy and correcting Love numbers $k_2$.

# Acknowledgements

This study is supported by National Science and Technology Major Project for Deep Earth Research (Grant 2024ZD1002700), the Central-to-Local Science and Technology Development Special Project in Hubei Province (Grants 2025CFC006), the National Precise Gravity Measurement Facility (Huazhong University of Science and Technology) (Grant PGMF-2024-Z003), Theory of Hydrocarbon Enrichment under Multi-Spheric Interactions of the Earth (Grants THEMSIE). We are also grateful for the support from the Scuola di Ingegneria Aerspaziale of Sapienza University of Rome (Italy) , Lanzhou University and International Center for Theoretical Physics Asia-Pacific.

## Author contributions

The authors declare no conflict of interest. Xizhi Hu, under the guidance of Ignazio Ciufolini and Xiaodong Chen, defined the scientific objectives of this study and designed the specific research plan. Xiaodong Chen, Ignazio Ciufolini, Wei-Tou Ni, and Jianqiao Xu participated in the adaptation and optimization of the research methodology, as well as in-depth discussions of the results. Xizhi Hu implemented the programming and numerical computations based on established methods, supplemented new analytical results, integrated data from previous studies, and drafted the initial manuscript. Wei-Tou Ni, Antonio Paolozzi, Jianqiao Xu, and Xiaodong Chen critically reviewed and revised the manuscript for important intellectual content. All authors reviewed the manuscript and contributed individually to the writing.

## Funding

This work was funded by National Science and Technology Major Project for Deep Earth Research (Grant 2024ZD1002700); the Central-to-Local Science and Technology Development Special Project in Hubei Province (Grant 2025CFC006); the National Precise Gravity Measurement Facility (Huazhong University of Science and Technology) (Grant PGMF-2024-Z003); Theory of Hydrocarbon Enrichment under Multi-Spheric Interactions of the Earth (Grant THEMSIE).

## Data availability

The orbital element data of LARES 2 and LAGEOS used in this study are derived from satellite laser ranging (SLR) observations processed by the GEODYN II and UTOPIA orbit estimators (Ciufolini et al., 2023). The Earth tide constituent data are based on the Cartwright and Tayler (1971) convention. All numerical calculation results and key tidal perturbation parameters supporting the findings of this study are included in the published article and its supplementary tables. The raw SLR observation data are available from the corresponding author on reasonable request.